\documentclass[
  a4paper,
  fontsize=11pt,
  captions=tableheading,
  parskip=never,
  ]{scrartcl}

\setkomafont{caption}{\small}
\setkomafont{captionlabel}{\bfseries}
\addtokomafont{publishers}{\large}
\addtokomafont{subject}{\mdseries\large}

\pdfoutput=1

\usepackage{ILD}
\usepackage{graphicx}
\usepackage[symbol]{footmisc}
\usepackage{amssymb}
\usepackage{svg}

\usepackage[utf8]{inputenx}
\usepackage[T1]{fontenc}
\usepackage[british]{babel}
\usepackage{csquotes}

\usepackage{caption, subcaption}
\usepackage{import}
\usepackage{xspace}
\usepackage[shortcuts]{extdash}
\usepackage{siunitx}
\DeclareSIUnit \lightspeed {\text{{c}}}
\usepackage{xcolor}
\definecolor{linkblue}{HTML}{264772}
\usepackage{hyperref}
\hypersetup{
    colorlinks=true,
    linktocpage=true,
    linkcolor=linkblue,
    citecolor=linkblue,
    urlcolor=linkblue
}

\usepackage{wrapfig}

\graphicspath{ {./Pictures/} }
\DeclareGraphicsExtensions{.pdf,.png,.jpg}

\begin{document}

\hyphenation{
  am-pli-fi-ca-tion
  col-lab-o-ra-tion
  per-for-mance
  sat-u-rat-ed
  se-lect-ed
  spec-i-fied
}

\title{Towards an update of the ILD ZHH analysis}
\author{Bryan Bliewert, Jenny List, Dimitris Ntounis, Junping Tian, Julie Munch Torndal, Caterina Vernieri}

\titlecomment{Talk presented at the International Workshop on Future Linear Colliders (LCWS 2024), 8-11 July 2024, Tokyo. This work was carried out in the framework of the ILD Concept Group.}
\date{}

\addauthor{Bryan Bliewert}{\institute{1}\institute{2}}
\addauthor{Jenny List}{\institute{1}}
\addauthor{Dimitris Ntounis}{\institute{3}}
\addauthor{Junping Tian}{}
\addauthor{Julie Munch Torndal}{\institute{1}\institute{2}}
\addauthor{Caterina Vernieri}{\institute{3}}
\addinstitute{1}{Deutsches Elektronen\-/Synchrotron DESY, Germany}
\addinstitute{2}{Department of Physics, Universität Hamburg, Germany}
\addinstitute{3}{SLAC National Accelerator Laboratory, United States}
\addinstitute{4}{International Center for Elementary Particle Physics (ICEPP), The University of Tokyo, Japan}

\abstract{%
  The double Higgs-strahlungs process $e^+e^- \rightarrow ZHH$ allows to access the Higgs self-coupling at center-of-mass energies above $450$\,GeV. Its cross-section exhibits a very different behavior as a function of the value of the self-coupling than fusion-type processes like gluon-gluon fusion at LHC (and future hadron colliders) and $WW$ / $ZZ$ fusion at higher energy lepton colliders. Therefore it adds unique information to the picture, in particular should the value of the Higgs self-coupling differ from its Standard Model prediction. The last full evaluation of the potential of the ILC to measure this process is more than ten years old, and since then many of the reconstruction tools have received very significant improvements. This contribution presents the ongoing work in the ILD collaboration to update the ZHH projections for the next European Particle Physics Strategy Update.
}

\titlepage

\clearpage

\section{Introduction}
\label{sec:1Intro}

Since the discovery of a Standard Model (SM)-like Higgs boson at the LHC in 2012~\cite{ATLAS:2012yve, Bhat:2012zj}, measuring the value of the Higgs self-coupling $\lambda=\lambda_{hhh}$ to experimentally establish the mechanism of electroweak symmetry breaking is among the most important goals of particle physics.
\\\\
While in a pure tree-level picture deviations of $\lambda$ from its SM value are expected to be small if no deviations have been observed in other couplings, it has recently been shown~\cite{Arco2022-ou} that this is not at all true once loop corrections are included: Already in extension of the SM Higgs sector as simple as a Two-Higgs-Doublet-Model (2HDM), $\lambda$ can be driven to significantly larger values than in the SM by loop corrections with the additional Higgs bosons, \emph{even if all other couplings are perfectly SM-like}. Thus it is of utmost importance to plan future colliders such that they are able to measure $\lambda$ with the least possible model-dependence.
\\\\
Double-Higgs production at the LHC and at $e^+e^-$ colliders with $\sqrt{s} > 450$\,GeV gives direct tree-level sensitivity to the self-coupling. As opposed to many other measurements, the experimentally achievable precision at any collider depends significantly on the actual value of $\lambda$ realised in nature. A unique feature of the double Higgs-strahlungs process $e^+e^- \rightarrow ZHH$ is that its cross-section and the sensitivity to the value of $\lambda$ increase when $\lambda > \lambda_{\mathrm{SM}}$~\cite{DiMicco:2019ngk}. This holds in particular in the region where $\lambda$ is between 50\% and a few times larger than the SM prediction, which is of special interest to models of electroweak baryogenesis. The cross-sections and the resulting sensitivity of vector--boson--fusion--like processes, including both $gg$ fusion in $pp$ collisions as well as $WW$ fusion in $e^+e^-$ collisions at higher center-of-mass energies, instead drops when $\lambda > \lambda_{\mathrm{SM}}$.
over a wide range of potential value of $\lambda$.
\\\\
This is a strong motivation for future $e^+e^-$ colliders which allow direct measurements of $\lambda$ from double-Higgs production. One of them is the International Linear Collider (ILC) with the International Large Detector (ILD) detector concept. An energy stage of this machine at an energy of about $500$\,GeV would allow access to double-Higgs production, and in particular to double Higgs-strahlung $e^+e^- \rightarrow ZHH$. The last full analysis of this process was carried out ten years ago~\cite{Duerig2016}. In light of the imminent input to the European Particle Physics Strategy Update (EPPSU) next year, there are now a renewed interest and a global effort to extract limits with state-of-the-art tools, benefiting both reconstruction and analysis. This effort aims to update the ILD projections for the $ZHH$ process, and the extraction of $\lambda$ based on detailed, Geant4-based detector simulation, including all relevant backgrounds. Also, following previous and promising studies~\cite{Torndal2024-ay}, an additional analysis at a slightly higher center-of-mass energy of $\sqrt{s}=\SI{550}{\giga\eV}$ will be carried out. Due to limited person power, the main focus will be on the $HH \rightarrow 4b$ channel, which covers $33.7\%$ of all double-Higgs events. For the $Z$ boson, the decays into jets, muons, electrons and neutrinos will be included, like in the previous analysis. Optionally the $Z$ decays into $\tau$ leptons could be added.
\\\\
The first section of this contribution presents improvements due to state-of-the-art tools developed during the last analysis, many of which are driven by machine-learning (ML). After that, pathways for further potential improvements are presented and a conclusion is given.

\section{State-of-the-art Analysis Tools}
\label{sec:2_AnalysisSOTA}

Since the last full $ZHH$ analysis, significant improvements could be made to reconstruction and analysis tools. The current benchmark for both jet clustering and tagging at ILD is LCFIPlus~\cite{Suehara16}. In the following, state-of-the-art methods for $b$-tagging, kinematic fitting and particle identification are presented.

\subsection{Jet-tagging with ParticleNet}
\label{subsec:2a_bTagging}

\begin{figure}[htbp]
    \centering
    \begin{subfigure}{.5\textwidth}
        \centering
        \includegraphics[width=0.95\textwidth]{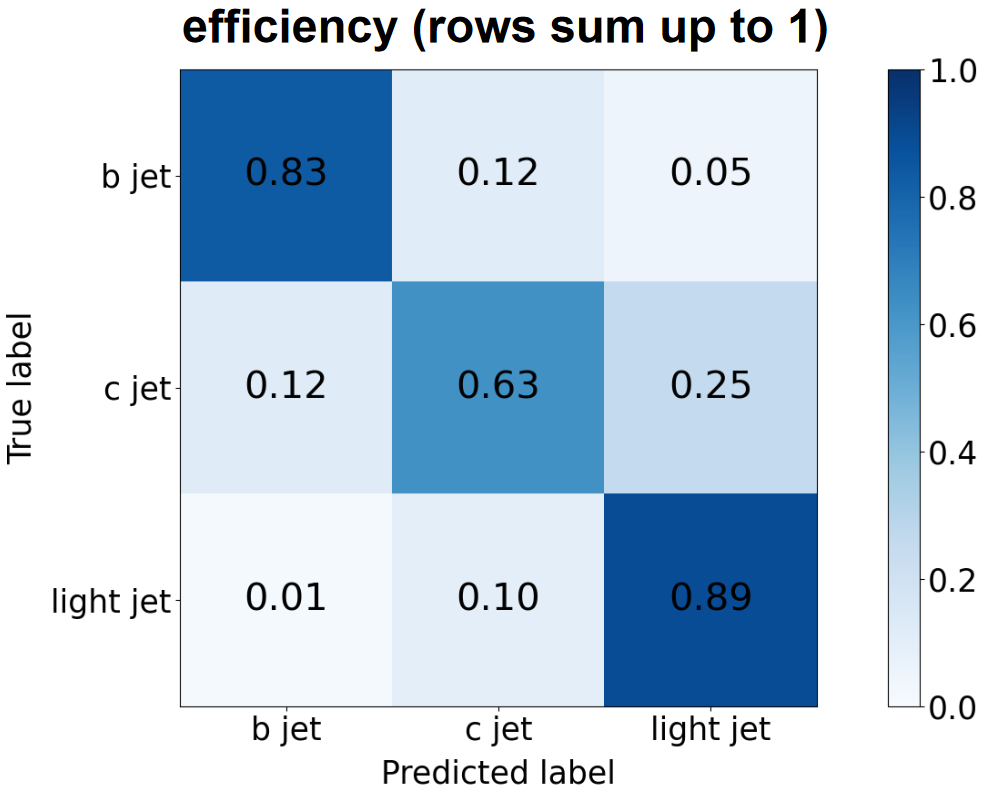}
        \caption{Jet-tagging efficiency}
        \label{fig:2a_bTaggingEff}
    \end{subfigure}\hfill%
    \begin{subfigure}{.5\textwidth}
        \centering
        \includegraphics[width=0.95\textwidth]{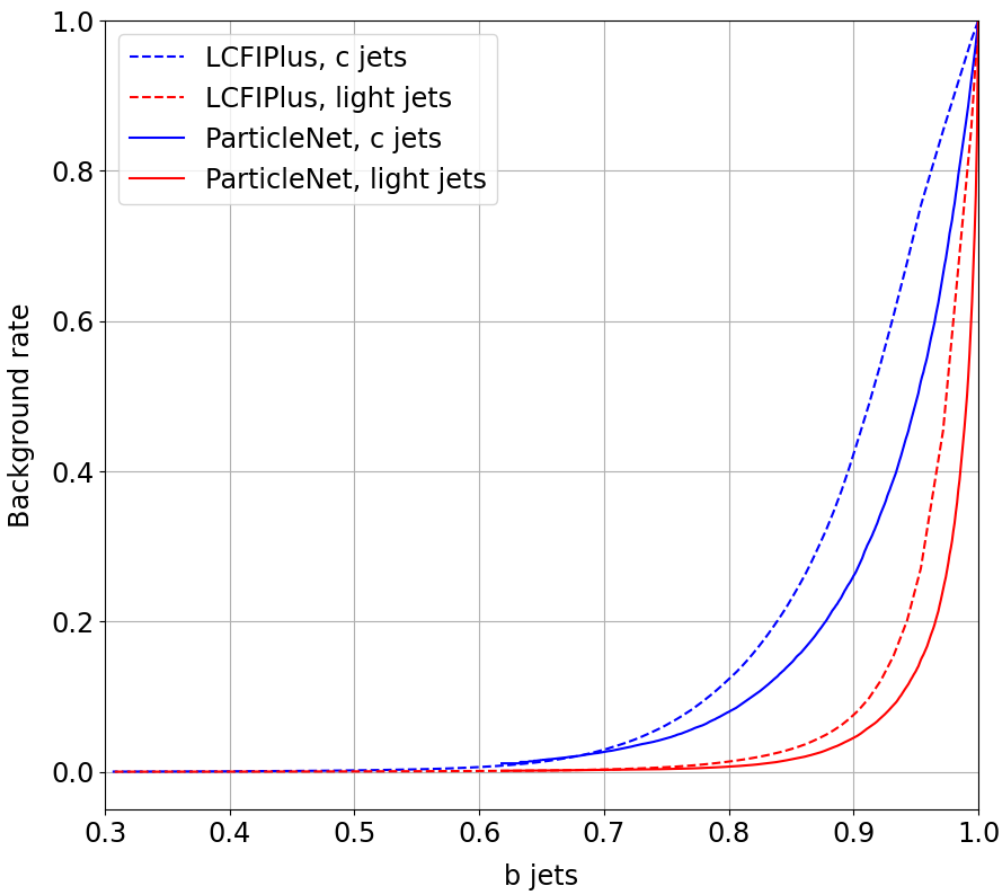}
        \caption{ROC-curve}
        \label{fig:2a_bTaggingROC}
    \end{subfigure}%
    
    \caption{Jet-tagging performance with ParticleNet at ILD and ILC500. (a) Tagging efficiencies for separating $b$- from $c$- and light jets at an example working point. (b) Performance comparison of $b$- vs.\ $c$- and light-jet tagging by ROC curves with the benchmark algorithm LCFIPlus. The plot shows background efficiency vs.\ signal efficiency (the closer to the bottom-right corner, the better). From~\cite{Meyer23}.}
    \label{fig:2a_bTagging}
\end{figure}

One crucial parameter in the $ZHH$ analysis is the $b$-tagging efficiency. In the $4b$-channel, improvements of the $b$-tagging efficiency (at the same background rejection rates) enter the analysis by forth power.
\\\\
In recent years, machine learning could be applied very successfully~\cite{Huilin20,Huilin22} to the task of jet-flavour tagging. One example is the ParticleNet architecture, which treats jets as point clouds and is thus permutation invariant. It is able to process feature data of both jet constituents (i.e.\ particle flow objects, PFOs) and secondary vertices. The latter is beneficial for the tagging of heavy quark jets such as $b$-jets. ParticleNet has been implemented~\cite{Meyer23} in the Marlin framework and trained using fully-simulated physics samples with $6$ jets at ILD with a center-of-mass energy of $\sqrt{s}=\SI{500}{\giga\eV}$. A confusion matrix including $b$, $c$ and light ($u,d,s$) quarks as well as a ROC-curve comparing the ParticleNet model with LCFIPlus are shown in Fig.~\ref{fig:2a_bTagging}.

As can be seen in Fig.~\ref{fig:2a_bTaggingEff}, a $b$-tagging efficiency $\epsilon_{b}$ of around $83\%$ could be obtained by using the ParticleNet model at an example working point with a mis-tag rate of $12\%$ to falsely identify a $c$ jet as $b$ jet. Comparing the overall performance of the ML-based tagger with that of LCFIPlus as in Fig.~\ref{fig:2a_bTaggingROC} shows a consistently superior background suppression when identifying $b$-jets. For the separation of $b$ jets with light jets ($c$ jets), an up to 4\% increase in efficiency over the already improved LCFIPlus~\cite{Suehara17} can be gained at the same purities, which are around 99\% (88\%). Earlier studies \cite{Duerig2016} have shown that an up to 11\% relative improvement on the precision of the self-coupling can be expected from a 5\% relative increase in $\epsilon_b$. The  results in Fig.~\ref{fig:2a_bTagging} show that this expected improvement is clearly within reach.

\subsection{Kinematic fitting}
\label{subsec:2b_KinFit}

\begin{figure}[htbp]
    \centering
    \includegraphics[width=0.5\textwidth]{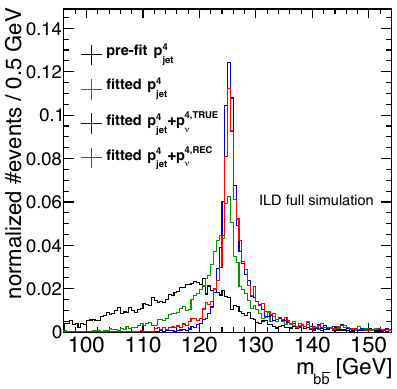}
    
    \caption{ Reconstructed di-jet masses from $H \rightarrow b\bar{b}$ decays before and after kinematic fitting: no correction (black), kinematic fitting without (green), with reconstructed (red) and with cheated (blue) neutrino momenta. From~\cite{Radkhorrami23}. }
    \label{fig:2b_kinfitDijetHiggs}
\end{figure}

Another important ingredient to the previous $ZHH$ analysis was kinematic fitting.
At the time of the previous analysis, the kinematic fit already was able to account for initial-state photon radiation~\cite{Beckmann:2010ib}, but the correction for missing energy from semi-leptonic heavy flavour decays inside the $b$ jets as well as the overall estimate of the measurement errors on the jet energy and the jet angles were very rudimentary. In both areas, significant progress has been achieved in the last years~\cite{Einhaus:2022bnv, Radkhorrami23, Radkhorrami25}, leading to a massive improvement of the kinematic reconstruction of $H\rightarrow b\bar{b}$ decays, illustrated in Fig.~\ref{fig:2b_kinfitDijetHiggs} for the somewhat simpler case of $ZH \rightarrow \mu^+\mu^- b\bar{b}$ at $\sqrt{s}=250$\,GeV.

Transferring these developments to the more complex case of $ZHH$ events is work-in-progress, which is expected to help both with identifying the correct di-jet pairs originating from one boson as well as to separate $ZHH$ events from the most challenging backgrounds like $ZZH$ production.

\subsection{Comprehensive Particle Identification (CPID)}
\label{subsec:2c_CPID}

\begin{figure}[htbp]
    \centering
    \begin{subfigure}{.5\textwidth}
        \centering
        \includegraphics[width=0.95\textwidth]{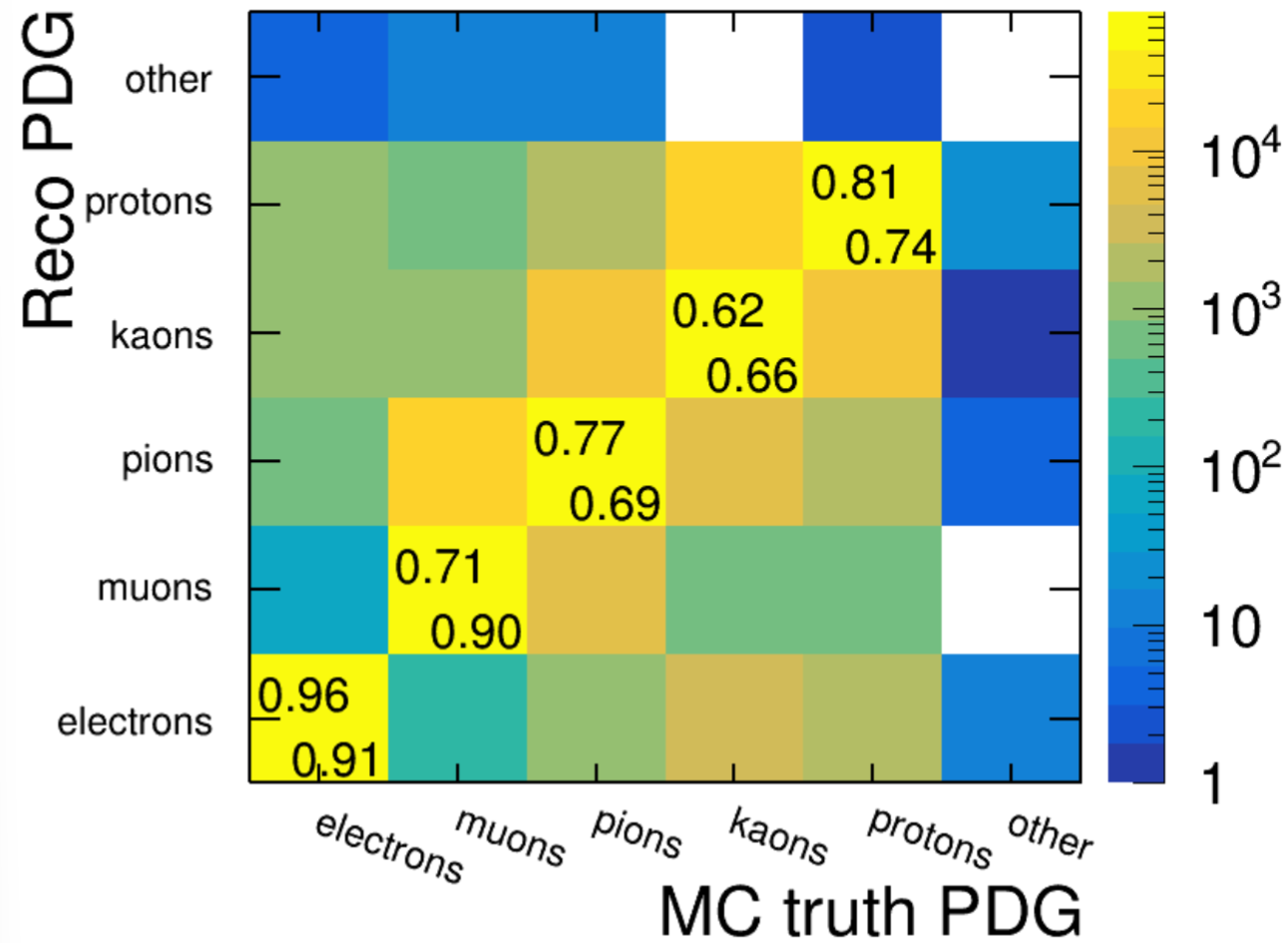}
        \caption{Likelihood PID}
        \label{fig:2c_PIDconfusionLikelihoodPID}
    \end{subfigure}\hfill%
    \begin{subfigure}{.5\textwidth}
        \centering
        \includegraphics[width=0.95\textwidth]{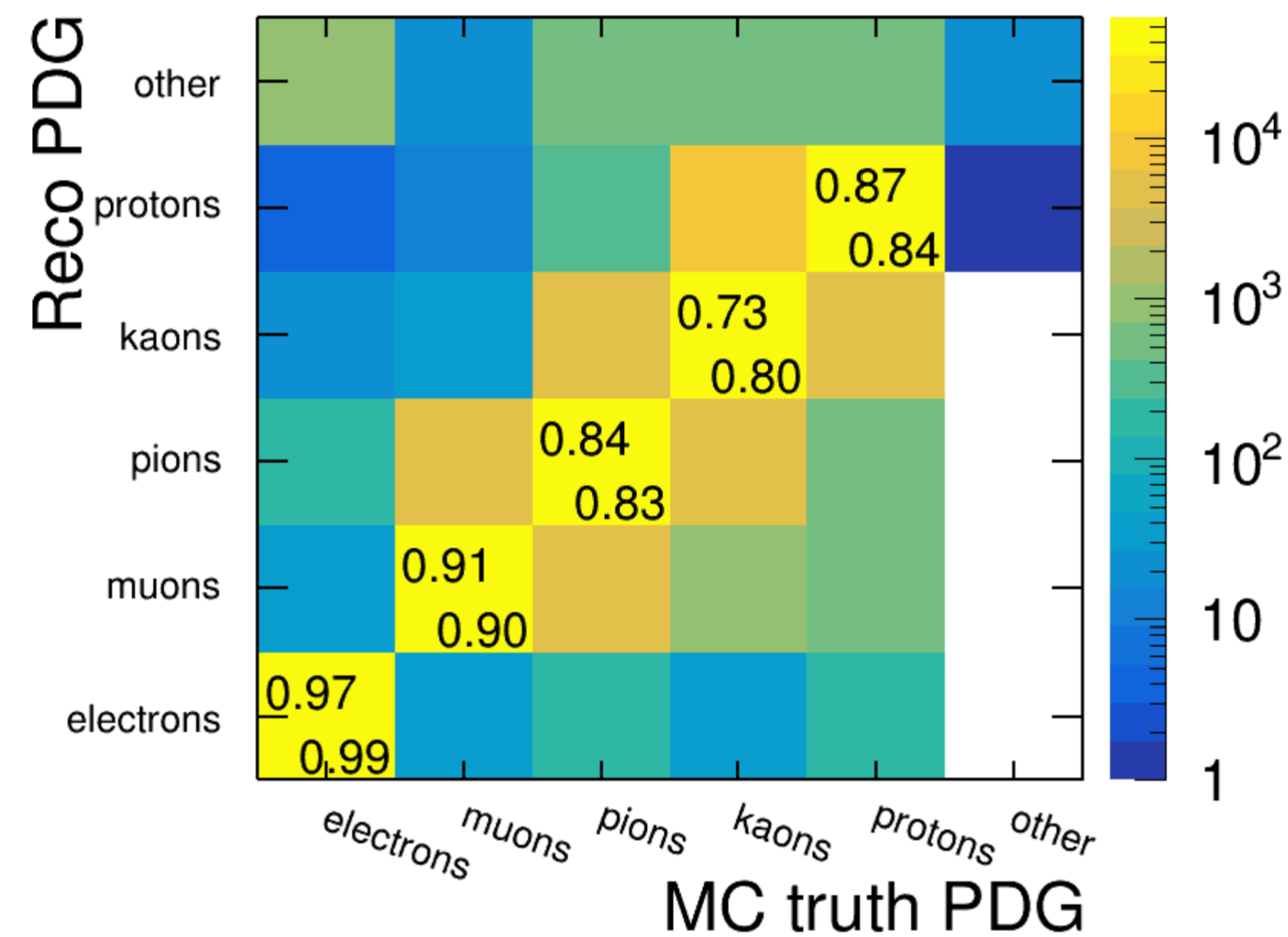}
        \caption{CPID}
        \label{fig:2c_PIDconfusionCPID}
    \end{subfigure}%
    
    \caption{Confusion matrices showing the performance of (a) the benchmark PID algorithm and (b) the new CPID (right). On the diagonal elements, efficiency and purity are given in the top left and bottom right, respectively. From~\cite{Einhaus23}. }
    \label{fig:2c_PIDconfusion}
\end{figure}

The $ZHH$ analysis profits further from improvements in particle identification (PID) in multiple ways. One example is the leptonic signal channel. There, after an isolated lepton criterion selects charged leptons, only leptons of the same family and with opposite charge are included in the pairing to reconstruct the leptonic $Z$ decay. Because two leptons are paired, improvements to the PID efficiency of electrons and muons go into the analysis squared.
\\\\
As a PID framework, CPID~\cite{Einhaus23} has been designed from the ground-up towards modularity for both different inference modules as well as data sources and detector systems. These range over $dE/dx$, cluster shape and time-of-flight (TOF). The latter has so far not been included in the benchmark PID solution (``LikelihoodPI''). An example of the performance gain when using CPID at ILD is illustrated in Fig.~\ref{fig:2c_PIDconfusion}.

The confusion matrix for CPID in Fig.~\ref{fig:2c_PIDconfusionCPID} shows excellent separation for leptons and still very good results for separating pions, kaons and protons. Compared to the benchmark, the PID efficiency is significantly improved by a relative 10-25\% for all considered particles except electrons. For all particles except the muons, also the purity is improved by a relative 15-20\% at the same time.

\section{Future Analysis Tools}
\label{sec:3_Analysis2Future}

Additional to the state-of-the-art and production-ready improvements described earlier, two challenges for future research shall be presented here with proof-of-concept solutions. They address the tasks of jet clustering and event classification.  

\subsection{More accurate jet clustering with ML}
\label{subsec:3a_JetClustering}

Previous studies~\cite{Duerig2016} have shown that the jet clustering is one of the leading sources of error in the $ZHH$ analysis, impacting the sensitivity to $\lambda$ by approximately a factor of two. This is due to the interplay of two effects: mis-clustering and mis-pairing of jets to di-jets. They can be illustrated by making use of a unique feature of the event data model~\cite{Gaede03}, namely mappings between particles and di-jets from reconstruction and their corresponding counterpart from full simulation. Using these mappings of PFOs to di-jets and matching each reconstructed di-jet to it's ``true counterpart'' by angular overlap, a ``true'' di-jet can be defined and its associated mass calculated. 

\begin{figure}[htbp]
    \centering
    \includegraphics[width=0.5\textwidth]{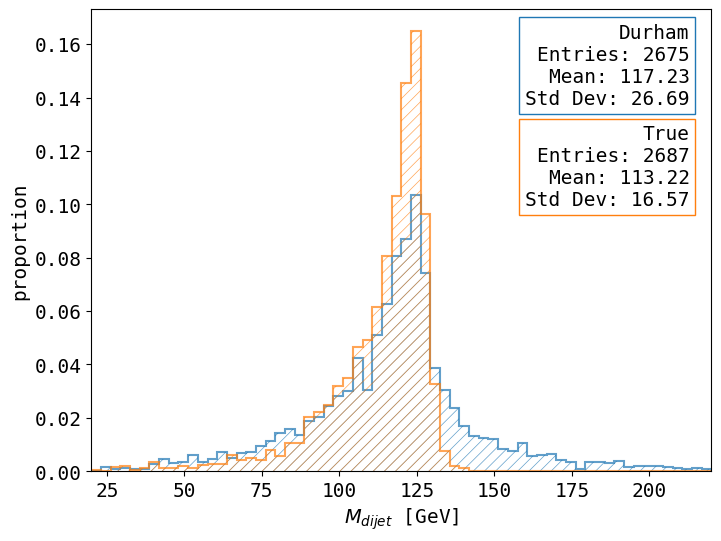}
    
    \caption{Reconstructed di-jet masses with the Durham algorithm and cheated (``True'') using TrueJet~\cite{Berggren18}. }
    \label{fig:3a_JetClusteringTrue}
\end{figure}

This is shown for hadronically decaying Higgs bosons in $ZHH$ events in Fig.~\ref{fig:3a_JetClusteringTrue}. The much sharper mass distribution for the ``true'' di-jets shows that a more accurate jet clustering algorithm would significantly benefit any downstream analysis that uses the reconstructed di-jet mass. For the Durham case, the jets are paired to two di-jets using the best $\chi^2$ matching assuming the Higgs mass.
\\\\
A more thorough analysis is possible by quantifying the purity $\pi_E$ and efficiency $\epsilon_E$ of each di-jet's energy~\cite{Torndal2024-ay}. When imposing a 95\% boundary on both axes, all di-jets can be grouped into four categories as given in Table~\ref{table:3a_jetCategorization}.

\begin{table}[htbp]
    \centering
    \begin{tabular}{ c|c|c }
        Category & $\pi_E = E_{corr} / E_{true}$ & $\epsilon_E = E_{corr} / E_{reco}$ \\
        \hline
        A & $\geq 95\%$ & $\geq 95\%$ \\
        B & $\geq 95\%$ & $< 95\%$ \\
        C & $< 95\%$ & $\geq 95\%$ \\
        D & $< 95\%$ & $< 95\%$ \\
    \end{tabular}
    
    \caption{Categorization of reconstructed di-jets by energy fraction. $E_{corr}$ is the energy of the correctly clustered PFOs (i.e.\ union of PFOs present in the reconstructed and associated true di-jet).}
    \label{table:3a_jetCategorization} 
\end{table}

\begin{figure}[htbp]
    \centering
    \begin{subfigure}{.5\textwidth}
        \centering
        \includegraphics[width=0.95\textwidth]{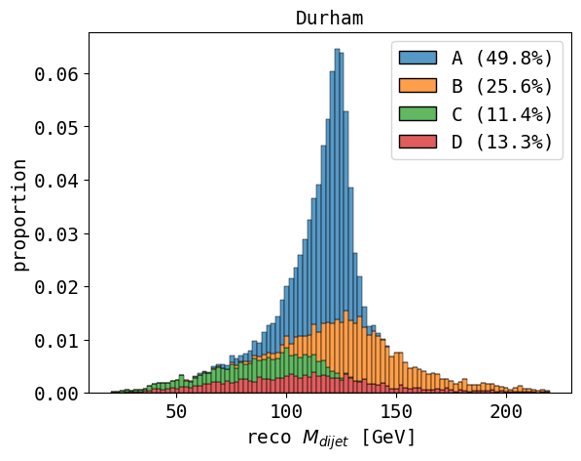}
        \caption{Durham}
        \label{fig:3a_clusteringDurham}
    \end{subfigure}\hfill%
    \begin{subfigure}{.5\textwidth}
        \centering
        \includegraphics[width=0.95\textwidth]{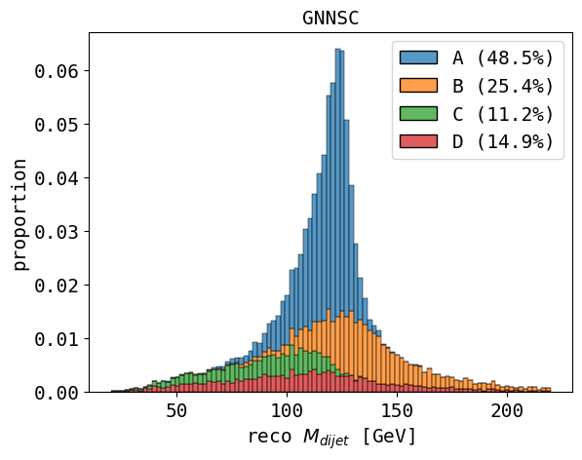}
        \caption{GNNSC}
        \label{fig:3a_clusteringGNNSC}
    \end{subfigure}%
    
    \caption{Di-jet masses for hadronically decaying Higgs bosons in ZHH samples at $\SI{500}{\giga\eV}$ as reconstructed by (a) the Durham algorithm and (b) the experimental GNNSC model.}
    \label{fig:3a_clustering}
\end{figure}

Assuming the categorization given in Table~\ref{table:3a_jetCategorization}, the higher the fraction of di-jets in category A and the lower in the other fractions, the better the average reconstruction performance.
\\\\
One possible approach towards more accurate jet clustering algorithms is to use ML, precisely graph neural networks (GNNs) to calculate scores for whether or not jet constituents shall be clustered together. Following this idea, a hybrid model (``GNNSC'') has been developed as a proof-of-concept study. This model uses the TransformerConv operator~\cite{Shi21} sandwiched between linear input layers and a Tanh activation layer to calculate scores, and spectral clustering (SC) on the affinity matrix $A \in \mathbb{R}^{n \times n}$ to create $m$ clusters for $n$ PFOs. The scoring module takes in only the four-momenta of the PFOs. It is trained fully-supervised on ZHH events and using labels from the true jet to PFO-matching described above (ignoring gluon splittings), while the spectral clustering requires no training. Results for reconstructed di-jet masses using Durham and GNNSC are shown in Fig.~\ref{fig:3a_clustering}.

The Higgs peak at $\SI{125}{\giga\eV}$ is well reconstructed using either Durham or the GNNSC method, as evident from Fig.~\ref{fig:3a_clustering}. More precisely, both the full-width at half-maximum (FWHM) of the two distributions and the distribution of the mis-clustering categories A-D over the di-jets are effectively indistinguishable. Further studies have shown that the model successfully generalizes to other physics processes (e.g.\ $ZZH$) well. This leads us to the conclusion that already with reasonably complicated ML architectures, it is possible to reach the performance of current benchmark jet clustering algorithms. More refined architectures with inbuilt physical symmetries and properties (IRC-safety) as well as fully-differentiable models could allow additional gains and allow to surpass the performance of current jet clustering algorithms at future Higgs factories.

\subsection{Event classification with Matrix Elements}
\label{subsec:3b_MEM}

\begin{figure}[htbp]
    \centering
    \begin{subfigure}{.5\textwidth}
        \centering
        \includegraphics[width=0.5\textwidth]{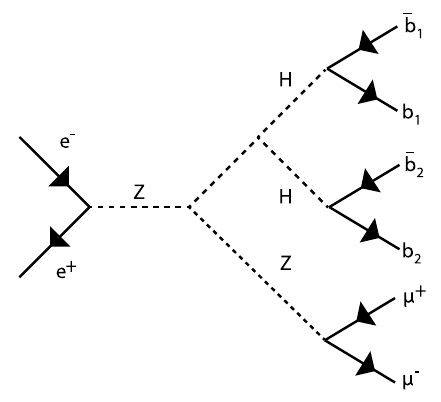}

        \label{fig:3b_feynmanZHH}
    \end{subfigure}\hfill%
    \begin{subfigure}{.5\textwidth}
        \centering
        \includegraphics[width=0.5\textwidth]{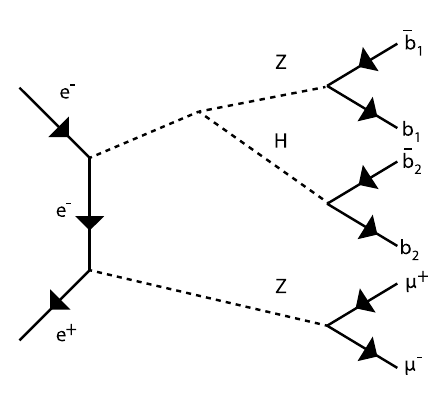}

        \label{fig:3b_feynmanZZH}
    \end{subfigure}%
    
    \caption{Some of the Feynman diagrams for (a) the $ZHH$ and (b) the $ZZH$ process.}
    \label{fig:3b_feynman}
\end{figure}

An important challenge in the $ZHH$ analysis is the separation of irreducible backgrounds, for example $ZZH$ events. Example Feynman diagrams for both processes in the leptonic signal channel (here $Z \rightarrow \mu^+\mu^-$) are given in Fig.~\ref{fig:3b_feynman}.

\begin{figure}[htbp]
    \centering
    \includegraphics[page=1,width=.6\textwidth,trim = .2cm .4cm .2cm 1.8cm, clip]{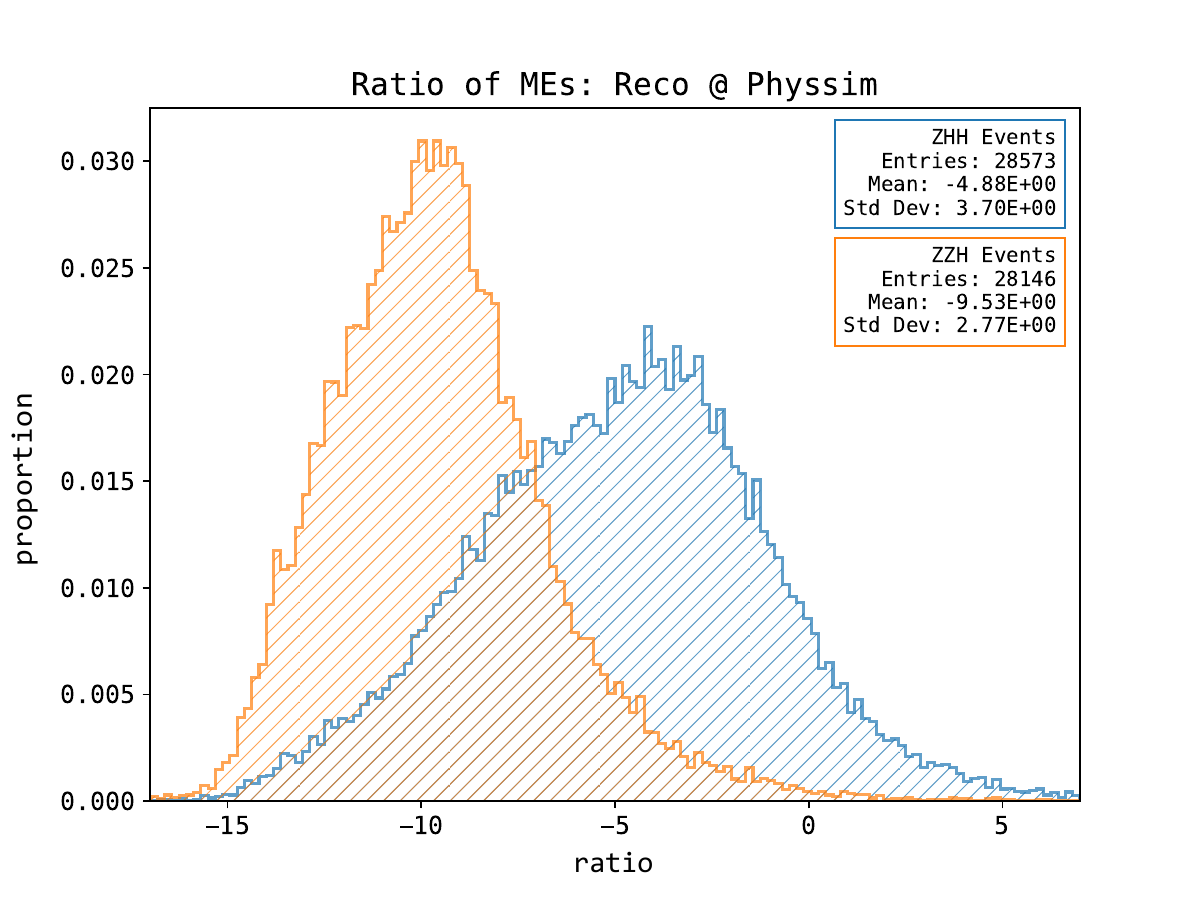}
    \caption{ Ratio of matrix elements for the $ZHH$ and $ZZH$ hypothesis. }
    \label{fig:3b_meRecoRaw}
\end{figure}

Detector effects as well as missing momentum from semi-leptonic decays of B-mesons and other particles complicate the separation of $ZHH$ and $ZZH$ events when observables such as the invariant mass are used.
\\\\
Another observable for separating these processes and using all available kinematic information is the hard-scattering matrix element. It can be calculated for both processes and including the remaining contributing Feynman diagrams with the reconstructed kinematics of the final state leptons and $b$-jets. The ratio $r=M_{ZHH}/M_{ZZH}$ of the matrix elements can then be used to construct a discriminator. An example for a sample of $ZHH$ and $ZZH$ events is given in Fig.~\ref{fig:3b_meRecoRaw}.

\begin{figure}[htbp]
    \centering
    \includegraphics[page=2,width=.6\textwidth,trim = .2cm 0cm .2cm .5cm, clip]{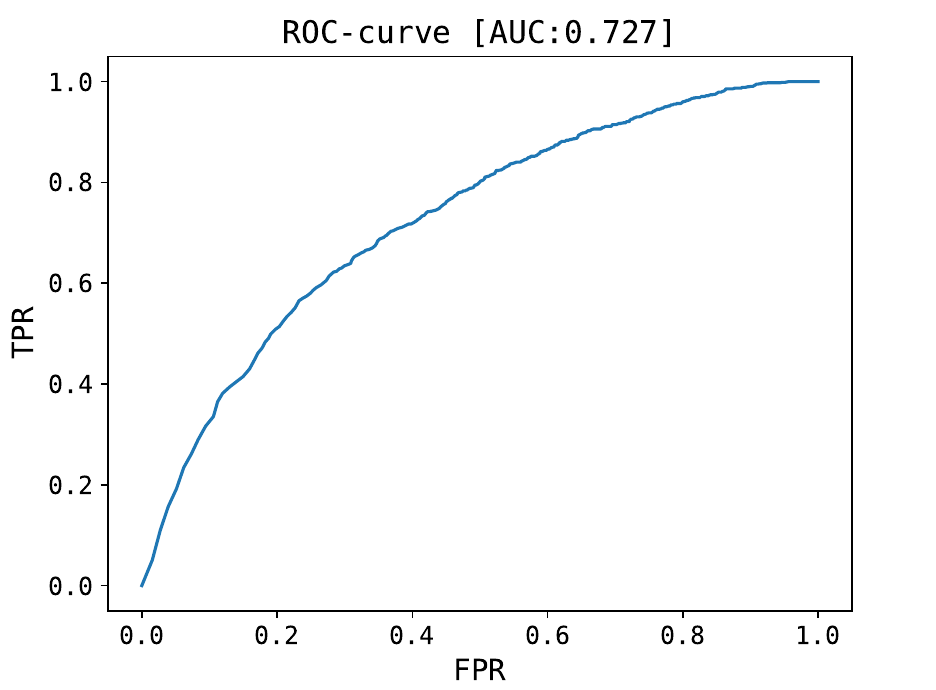}
    \caption{ ROC-curve of the $ZHH$/$ZZH$ discriminator using the full MEM procedure in terms of the true positive rate (TPR), i.e.\ signal efficiency, vs.\ false positive rate (FPR), i.e.\ background efficiency.}
    \label{fig:3b_meRecoFullVEGAS}
\end{figure}

The histogram in Fig.~\ref{fig:3b_meRecoRaw} shows that the hard-scattering matrix elements calculated with the reconstructed kinematics already provide much separation power. Here, the kinematics of the Higgs boson are reconstructed by adding the four momenta of the associated jets and for the $ZZH$ hypothesis, the average over the possible permutations is calculated. 
\\\\
However, to account for detector effects and parton showering, the approach must be extended. This procedure is known as the matrix element method (MEM) and described in literature~\cite{Kondo88,Volobouev11}. It allows to calculate the likelihood for an event to have happened under a certain hypothesis and is based on first principles. Instead of an evaluation at the measured kinematics, it requires a phase space integration over the matrix elements, functions describing the transfer of parton to jet kinematics and other features. As a proof-of-concept, this procedure has been carried with some simplifications, for example assuming perfect detector acceptance as well as efficiency and perfect measurement of the muon kinematics. Also, initial state radiation (ISR) is not accounted for. An on-shell condition has been assumed for the b quarks, such that the jet momentum transfer can be parameterized in spherical coordinates and using the energy and fixed mass. The transfer of jet angles and energy is manually fitted using Gaussian functions and the mappings between generator- and reconstructed level quanta (ignoring gluon splittings). Finally, the integral is solved using the VEGAS+ algorithm~\cite{Lepage21}. In accordance to the Neyman-Pearsson lemma, the two hypotheses can then be separated via the likelihood-ratio. The performance of the resulting discriminator is shown with a ROC-curve in Fig.~\ref{fig:3b_meRecoFullVEGAS}.

The discriminator shows acceptable performance in Fig.~\ref{fig:3b_meRecoFullVEGAS} with an area-under-curve (AUC) of $0.73$. In the future, improvements could be made by modeling the detector transfer functions more accurately (e.g. using normalizing flows) and developing dedicated phase space parameterizations for each process.

\section{Conclusion}
\label{sec:4_Conclusion}
About ten years ago, the expected precision on the Higgs self-coupling at ILC at $500$\,GeV has been evaluated based on detailed, Geant4-based simulation of the ILD detector concept. Based on the $ZHH$ process with $Z \rightarrow q\bar{q}, e^+e^-, \mu^+\mu^-, \nu \bar{\nu}$ and $HH \rightarrow b\bar{b}b\bar{b}/b\bar{b}WW^*$, and assuming the SM value of the self-coupling, the projection achieved a precision of $27\%$.
\\\\
The ongoing update of the $ZHH$ analysis will benefit from a variety of production-ready improvements, especially in jet tagging, PID and using kinematic fitting. We expect these to increase the sensitivity $\Delta\lambda_{SM}/\lambda_{SM}$ from $27\%$ to better than $20\%$. More improvements are possible in the removal of the $\gamma\gamma$-overlay (up to relative 15\%) and by including $Z \rightarrow \tau^+\tau^-$ in the leptonic channel (up to relative 8\%). This does neither include further improvements from a slightly higher $\sqrt{s}=\SI{550}{\giga\eV}$ nor from the inclusion of the $WW$-fusion production, which are both under study as well.
\\\\
Further future analysis improvements could be accomplished by developing more accurate jet clustering and event classification algorithms. Machine learning offers many opportunities for these tasks and can profit from more physical observables as well.

\section{Acknowledgments}
\label{sec:5_Acknowledgments}

We would like to thank the LCC generator working group and the ILD software working group for providing the simulation and reconstruction tools and producing the Monte Carlo samples used in this study. This work has benefited from computing services provided by the ILC Virtual Organization, supported by the national resource providers of the EGI Federation and the Open Science GRID. In this study we widely used the National Analysis Facility (NAF) \cite{Haupt2010-ct} and the Grid computational resources operated at Deutsches Elektronen-Synchrotron (DESY), Hamburg, Germany. We thankfully acknowledge the support by the Deutsche Forschungsgemeinschaft (DFG, German Research Foundation) under Germany’s Excellence Strategy EXC 2121 "Quantum Universe" 390833306.

\bibliographystyle{abbrv_mod}
\bibliography{sources}

\begin{thebibliography}{10}

\bibitem{ATLAS:2012yve}
G.~Aad et~al.
\newblock {\em {Observation of a new particle in the search for the Standard
  Model Higgs boson with the ATLAS detector at the LHC}}.
\newblock Phys. Lett. B, 716:1--29, 2012.
\newblock \url{http://dx.doi.org/10.1016/j.physletb.2012.08.020}.

\bibitem{Bhat:2012zj}
P.~C. Bhat.
\newblock {\em {Observation of a Higgs-like Boson in CMS at the LHC}}.
\newblock Nucl. Phys. B Proc. Suppl., 234:7--14, 2013.
\newblock \url{http://dx.doi.org/10.1016/j.nuclphysbps.2012.11.003}.

\bibitem{Arco2022-ou}
F.~Arco, S.~Heinemeyer, and M.~J. Herrero.
\newblock {\em Triple Higgs couplings in the {2HDM}: the complete picture}.
\newblock Eur. Phys. J. C Part. Fields, 82(6), 6 2022.
\newblock \url{http://dx.doi.org/10.1140/epjc/s10052-022-10485-9}.

\bibitem{DiMicco:2019ngk}
J.~Alison et~al.
\newblock {\em {Higgs boson potential at colliders: Status and perspectives}}.
\newblock Rev. Phys., 5:100045, 2020.
\newblock \url{http://dx.doi.org/10.1016/j.revip.2020.100045}.

\bibitem{Duerig2016}
C.~F. Duerig.
\newblock {\em Measuring the Higgs self-coupling at the international linear
  collider}.
\newblock Deutsches Elektronen-Synchrotron, DESY, Hamburg, 12 2016.

\bibitem{Torndal2024-ay}
J.~Torndal, J.~List, D.~Ntounis, and C.~Vernieri.
\newblock Higgs self-coupling measurement at future e+e- colliders.
\newblock In {\em Proceedings of The European Physical Society Conference on
  High Energy Physics --- {PoS(EPS-HEP2023})}, Trieste, Italy, 2 2024. Sissa
  Medialab.
\newblock \url{http://dx.doi.org/10.22323/1.449.0406}.

\bibitem{Suehara16}
T.~Suehara and T.~Tanabe.
\newblock {\em LCFIPlus: A framework for jet analysis in linear collider
  studies}.
\newblock Nuclear Instruments and Methods in Physics Research Section A:
  Accelerators, Spectrometers, Detectors and Associated Equipment,
  808:109--116, 2016.
\newblock \url{http://dx.doi.org/10.1016/j.nima.2015.11.054}.

\bibitem{Huilin20}
H.~Qu and L.~Gouskos.
\newblock {\em Jet tagging via particle clouds}.
\newblock Phys. Rev. D, 101:056019, Mar 2020.
\newblock \url{http://dx.doi.org/10.1103/PhysRevD.101.056019}.

\bibitem{Huilin22}
H.~Qu, C.~Li, and S.~Qian.
\newblock {\em Particle Transformer for Jet Tagging}.
\newblock arXiv [hep-ph], 2022.
\newblock \url{http://dx.doi.org/10.48550/arXiv.2202.03772}.

\bibitem{Meyer23}
M.~Meyer.
\newblock Machine learning flavour tagging for future higgs factories.
\newblock Presented at the Second ECFA Workshop on e+e- Higgs/EW/Top Factories
  in Paestum (Salerno), 10 2023.

\bibitem{Suehara17}
T.~Suehara, T.~Tanabe, M.~Kurata, and J.~Strube.
\newblock Status of lcfiplus.
\newblock Presentation at ILD software and technical meeting at the Institut de
  Physique Nucléaire de Lyon, 4 2017.

\bibitem{Beckmann:2010ib}
M.~Beckmann, B.~List, and J.~List.
\newblock {\em {Treatment of Photon Radiation in Kinematic Fits at Future e+ e-
  Colliders}}.
\newblock Nucl. Instrum. Meth. A, 624:184--191, 2010.
\newblock \url{http://dx.doi.org/10.1016/j.nima.2010.08.107}.

\bibitem{Einhaus:2022bnv}
U.~Einhaus, B.~Dudar, J.~List, Y.~Radkhorrami, and J.~Torndal.
\newblock {\em {Impact of Advances in Detector Techniques on Higgs Measurements
  at Future Higgs Factories}}.
\newblock PoS, ICHEP2022:538, 2022.
\newblock \url{http://dx.doi.org/10.22323/1.414.0538}.

\bibitem{Radkhorrami23}
Y.~Radkhorrami.
\newblock Reconstruction of heavy flavour jets for higgs physics.
\newblock Presented at the ILD Software and Analysis meeting, 6 2023.

\bibitem{Radkhorrami25}
Y.~Radkhorrami.
\newblock Reconstruction of heavy flavour jets for higgs physics.
\newblock PhD thesis in preparation, 2025.

\bibitem{Einhaus23}
U.~Einhaus.
\newblock {\em {CPID}: A Comprehensive Particle Identification Framework for
  Future e$^+$e$^-$ Colliders}, 8 2023.
\newblock \url{http://dx.doi.org/10.48550/arXiv.2307.15635}.

\bibitem{Gaede03}
F.~Gaede, T.~Behnke, N.~Graf, and T.~Johnson.
\newblock {\em LCIO - A persistency framework for linear collider simulation
  studies}, 2003.
\newblock \url{http://arxiv.org/abs/physics/0306114}.

\bibitem{Berggren18}
M.~Berggren.
\newblock Truth information : {RecoMCTruthLinker} and {TrueJet}.
\newblock Presented at the ILDsw meeting, KEK, 2 2018.

\bibitem{Shi21}
Y.~Shi, Z.~Huang, S.~Feng, H.~Zhong, W.~Wang, and Y.~Sun.
\newblock {\em Masked Label Prediction: Unified Message Passing Model for
  Semi-Supervised Classification}, 2021.
\newblock \url{http://arxiv.org/abs/2009.03509}.

\bibitem{Kondo88}
K.~Kondo.
\newblock {\em Dynamical Likelihood Method for Reconstruction of Events with
  Missing Momentum. I. Method and Toy Models}.
\newblock Journal of the Physical Society of Japan, 57(12):4126--4140, 1988.
\newblock \url{http://dx.doi.org/10.1143/JPSJ.57.4126}.

\bibitem{Volobouev11}
I.~Volobouev.
\newblock {\em Matrix Element Method in HEP: Transfer Functions, Efficiencies,
  and Likelihood Normalization}, 2011.
\newblock \url{http://arxiv.org/abs/1101.2259}.

\bibitem{Lepage21}
G.~P. Lepage.
\newblock {\em Adaptive multidimensional integration: vegas enhanced}.
\newblock Journal of Computational Physics, 439:110386, 2021.
\newblock \url{http://dx.doi.org/https://doi.org/10.1016/j.jcp.2021.110386}.

\bibitem{Haupt2010-ct}
A.~Haupt and Y.~Kemp.
\newblock {\em The {NAF}: National analysis facility at {DESY}}.
\newblock J. Phys. Conf. Ser., 219(5):052007, 4 2010.
\newblock \url{http://dx.doi.org/10.1088/1742-6596/219/5/052007}.

\end{thebibliography}

\end{document}